\pdfoutput=1
\documentclass[letterpaper, 10 pt, conference]{ieeeconf}  

\IEEEoverridecommandlockouts                              
\usepackage{cite}
\usepackage{graphicx} 
\usepackage{epsfig} 
\usepackage{mathptmx} 
\usepackage{times} 
\usepackage{amsmath} 
\usepackage{amssymb}  
\usepackage{xcolor} 
\usepackage{mathtools}
\usepackage{algorithm}
\usepackage{algpseudocode}
\usepackage{multirow} 
\usepackage{booktabs} 
\usepackage{subcaption} 

\title{\LARGE \bf
Phase Re-service in Reinforcement Learning Traffic Signal Control
}

\author{Zhiyao Zhang\textsuperscript{a, c}, George Gunter\textsuperscript{a, c}, Marcos Quinones-Grueiro\textsuperscript{c}, Yuhang Zhang\textsuperscript{a, c},  \\William Barbour\textsuperscript{c}, Gautam Biswas\textsuperscript{b, c}, and Daniel Work\textsuperscript{a, c}
\thanks{The authors are with \textsuperscript{a}Department of Civil and Environmental Engineering, \textsuperscript{b}Department of Computer Science, \textsuperscript{c}Institute for Software Integrated Systems, Vanderbilt University, USA. Email: zhiyao.zhang@vanderbilt.edu}%
\thanks{This work is supported by a grant from the U.S. Department of Transportation Grant Number 693JJ22140000Z44ATNREG3202. This material is based upon work supported by the National Science Foundation under Grant No. CNS-2135579 (Work). The contents of this article reflect the views of the authors, who are responsible for the facts and accuracy of the information presented herein. The U.S. Government assumes no liability for the contents or use thereof.}
\thanks{© 2024 IEEE.  Personal use of this material is permitted.  Permission from IEEE must be obtained for all other uses, in any current or future media, including reprinting/republishing this material for advertising or promotional purposes, creating new collective works, for resale or redistribution to servers or lists, or reuse of any copyrighted component of this work in other works.}
        }
\begin{document}

\maketitle

\begin{abstract}
This article proposes a novel approach to traffic signal control that combines phase re-service with reinforcement learning (RL). The RL agent directly determines the duration of the next phase in a pre-defined sequence. Before the RL agent's decision is executed, we use the shock wave theory to estimate queue expansion at the designated movement allowed for re-service and decide if phase re-service is necessary. If necessary, a temporary phase re-service is inserted before the next regular phase. We formulate the RL problem as a semi-Markov decision process (SMDP) and solve it with proximal policy optimization (PPO). We conducted a series of experiments that showed significant improvements thanks to the introduction of phase re-service. Vehicle delays are reduced by up to 29.95\% of the average and up to 59.21\% of the standard deviation. The number of stops is reduced by 26.05\% on average with 45.77\% less standard deviation.
\end{abstract}


\section{Introduction}
Dynamically changing traffic patterns is a core challenge in managing traffic at intersections, and when unaccounted for, they can
cause an increase in congestion and travel delay. \textit{Adaptive traffic signal control} (ATSC) offers a promising solution to mitigate congestion and enhance traffic flow. Significant ATSC deployments including SCOOT \cite{hunt1982scoot} and RHODES \cite{mirchandani2001rhodes} extract traffic patterns over time and optimize the signal timings accordingly. 

Recently, \textit{reinforcement learning} (RL) has emerged as a promising tool for ATSC, based on its learning and real-time computational capabilities in complex environments~\cite{wei2021recent}.  Though potentially promising algorithms have been developed (see~\cite{wei2021recent, noaeen2022reinforcement} for comprehensive surveys), phase starvation and safety concerns are known existing challenges to RL-based ATSC without fixed signal sequences~\cite{ibrokhimov2022biased}. Additionally, some algorithms experience a  performance drop in the presence of heavy traffic demands~\cite{zhang2023evaluation}.

 The left-turn movement is especially sensitive to high demand profiles because it often conflicts with oncoming traffic flow and has limited capacity. Performing phase re-service~\cite{stm}, which is when the controller  serves the same phase twice in one cycle, can be an effective approach to manage left-turn queue lengths \cite{wang2023leveraging}. Phase re-service has been successfully deployed in many real-world intersections~\cite{corey2012improving,lavrenz2015characterizing}.  While its implementation has been limited to traffic-responsive signal timing optimizations \cite{fang2006development}, we propose extending re-service to real-time ATSC for enhanced operational flexibility.
The main contribution of this article is that we introduce an approach to augment reinforcement learning-based adaptive traffic signal control to enable phase re-service.  The RL agent decides the duration of regular phases, and we use shock wave theory~\cite{stephanopoulos1979modelling, michalopoulos1981application,wang2017shockwave, cheng2012exploratory,liu2009real} to estimate queue growth and trigger phase re-service. Because the agent selects the phase duration, we model the control problem as a semi-Markov decision process (SMDP)~\cite{barto2003recent}. We test the performance of our approach on two intersection geometries, each with five demand profiles. The simulation results demonstrate that phase re-service significantly reduces vehicle delays and the number of stops overall, and for the protected left turn movement.


The remainder of this article is organized as follows: In Section \ref{background}, we provide the preliminaries for RL based signal control. Section \ref{formulation} presents the technical details of phase re-service determination, RL formulation and algorithm, and the pseudocode summarizing the training process. Experimental settings and result analysis are presented in Section \ref{experiments}. We finally conclude our work in Section \ref{conclusion}. 

\section{Preliminaries}\label{background}
\begin{figure}[htbp!]
    \centering
    \includegraphics[width=0.4\textwidth]{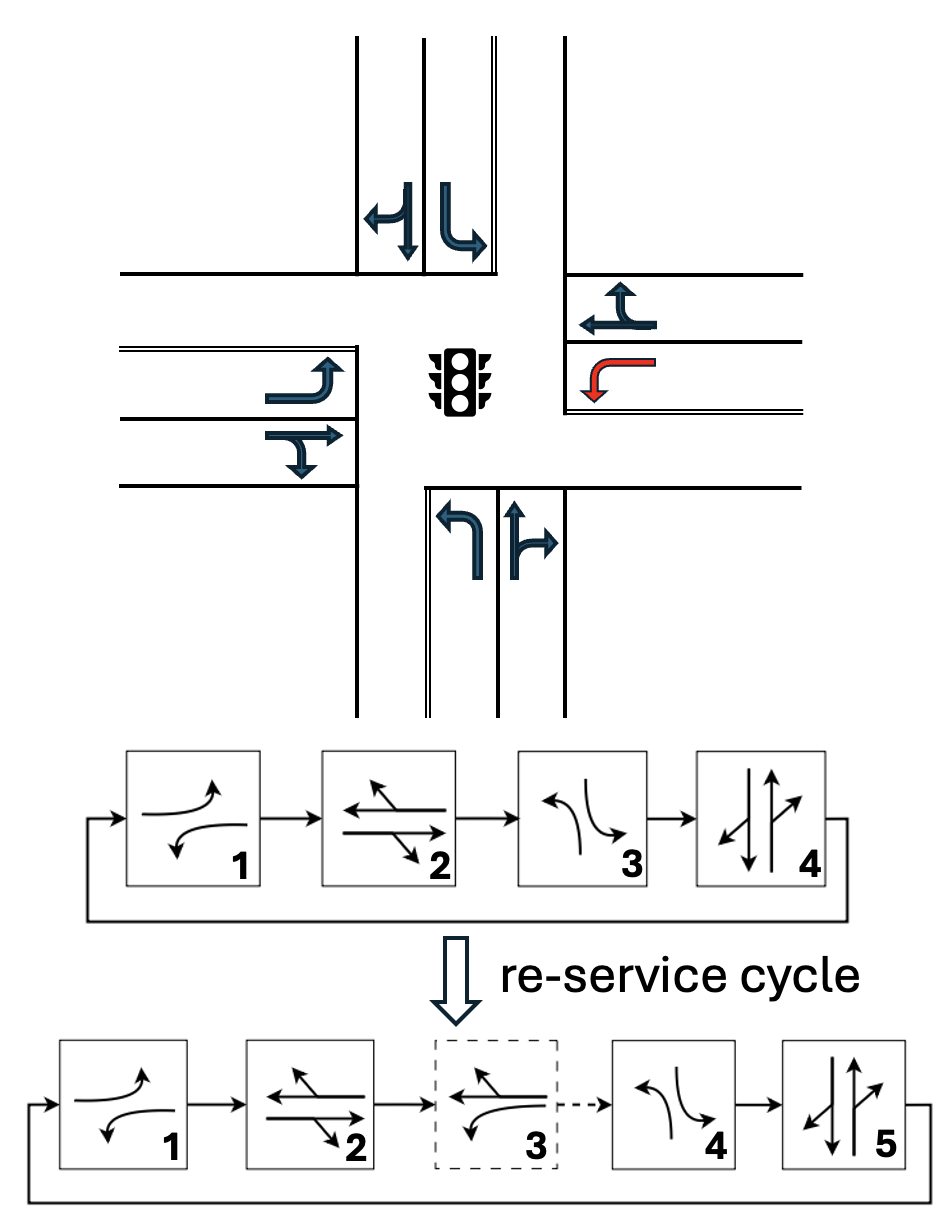}
    \caption{An example of intersection with a protected movement with phase re-service (red). In the top cycle, each phase is served once. In the bottom cycle, the high demand protected left turn movement is re-served in phase 3.}
    \label{fig:interchange}
\end{figure}


\subsection{Traffic signal control setup}
Consider traffic control at a single intersection with multiple incoming and outgoing roadways. Incoming vehicles travel on and are queued on incoming roadway \textit{lanes}. \textit{Queues} at signalized intersections are served by \textit{phases}, each of which groups one or more non-conflicting turning \textit{movements}. Phases are served with green signals in a pre-defined order known as the phase \textit{sequence}, and a complete iteration of the sequence is a \textit{cycle} \cite{stm}. 

At some intersections, particular left-turn movements face large demands during peak hours, such as the movement marked with red arrow in Fig. \ref{fig:interchange}. Serving a specific left-turn movement such as a protected left turn twice in a cycle can effectively clear excessive left-turn queues. The second service, known as \textit{phase re-service}, is typically pre-configured to follow the through movement in the same direction. For example, in Fig. \ref{fig:interchange}, the protected movement served in the first phase may be served again in the third phase. Our work considers adding phase re-service into RL based signal controllers.
\subsection{Semi-Markov decision processes} 
RL problems for signal control are often formulated as a \textit{Markov decision process} (MDP), $\{\mathit{S}, \mathit{A}, r, \mathit{P}, \gamma\}$. Here, $\mathit{S} \subseteq \mathbb{R}^{m} $ is the state space of the system in question with dimension $m$, $\mathit{A} \subseteq \mathbb{R}^{k}$ is the space of available actions with dimension $k$, $r: \mathit{S} \mapsto \mathbb{R}$ is a reward function, $\mathit{P}: \mathit{S} \times \mathit{A} \times \mathit{S} \mapsto \mathbb{R}$ is a probabilistic transition function, and $\gamma \in [0, 1]$ is a discount factor.

Additionally, let $s_{t} \in \mathit{S}$ be the system state at time $t$, and $a_{t} \in \mathit{A}$ be the action taken at time $t$. The actions are chosen at each step from a policy $a_{t} \sim \pi_{\theta}(a_{t}|s_{t})$ parameterized by $\theta$. RL attempts to maximize the long-term discounted reward $\sum_{i=0}^{\infty}\gamma^i r_i(\cdot)$ by choosing an optimal policy, typically through the selection of optimal policy parameters. Signal control problems in which the action at each timestep is the phase to serve are naturally written as MDPs, which can then be solved with RL. 



In this work we formulate a traffic signal control problem as a Semi-MDP, which is an extension of MDP which includes a varying transition time between states. The time between transitions is called the \textit{sojourn time}, denoted $j_{t}$, and we incorporate this into the state transition as $\mathit{P}(s_{t+1}, j_{t}|s_{t}, a_{t})$, meaning that the transition probability takes both the next state and the transition time to it into account. Signal control problems in which the action is the temporal duration to serve the next phase are naturally written as SMDPs.

The maximization of the long-term discounted reward is realized by converging the state value to Bellman optimality~\cite{barto2003recent,sutton2018reinforcement}, which for an SMDP is written as:
\begin{equation}
    V^*(s_t) = \max_{a \in A}\left [ r(s_t) + \sum_{s_{t+1}, j_t}\gamma^{j_t} \mathit{P}(s_{t+1}, j_t|s_t, a_t)V^*(s_{t+1}) \right].
\end{equation}



\subsection{Queue dynamics at intersections}

\begin{figure*}[tbp!]
    \centering
    \includegraphics[width=1\textwidth]{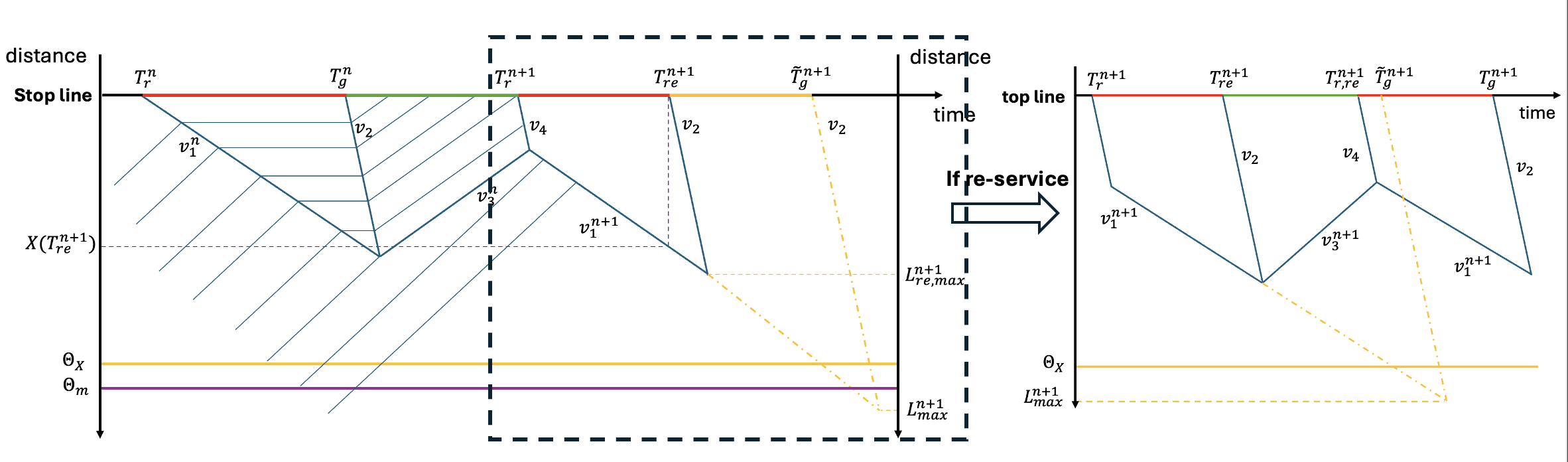}
    \caption{Demonstration of the shock wave at a single intersection lane over a signal cycle. Yellow transition is ignored for simplicity.}
    \label{fig:shock-wave-simple}
\end{figure*}

Let $n$ be an index on signal cycles. A set of important speed properties associated with shock waves at intersections is as follows~\cite{liu2009real, wang2017shockwave}:

\begin{align}
    v_1^n &= \frac{q_a^n}{k_j-k_a^n},\\
    v_2 &= \frac{q_m}{k_j-k_m},\\
    v_3^n &= \left|\frac{q_m-q_a^n}{k_m-k_a^n} \right|,\\
    v_4 &= \frac{-q_m}{k_m-k_j} = v_2,
\end{align}
where $k_m$, $k_j$, and $q_m$ are critical density, jam density, and saturation flow. Additionally, $k_a^n$ and $q_a^n$ are the density and flow for arriving vehicles at each cycle $n$. Modern sensors can directly measure the number of vehicles over a time window as the flow, and the average speed of these counted vehicles. Then the density is obtained from dividing the flow by the average vehicle speed.
Let $v_1^n$ be the queue expansion speed at the red signal, and let $v_3^n$ be the speed at which the end of queue moves forward to the stop line at the green signal, $v_2$ be the vehicle discharge speed from the queue at the green signal, and $v_4$ be the speed at which the residual of queued vehicles forms a new queue when the green signal ends (only for saturated movements). Fig.~\ref{fig:shock-wave-simple} shows graphically these different quantities.

\section{Controller design}~\label{formulation}
In order to handle intersections with high left-turn demand via ATSC, we propose an RL controller design approach combined with phase re-service based on queue length estimation from shockwave theory. We first frame the traffic control problem as an SMDP. Then, we describe how to determine the phase re-service from queue length estimation. Finally, we present a training algorithm.

\subsection{SMDP formulation}\label{smdp}

The traffic control problem can be formulated as an SMDP as follows:

\textbf{State} $s$. The state consists of the count of queuing vehicles in each incoming lane, the count of non-stopping vehicles in each incoming lane, as well as the number of phases until service. This information is commonly available from sensor data \cite{noaeen2022reinforcement}. 

\textbf{Action} $a$. The action taken by the RL agent is the duration of the next phase in a predefined cycle.  To account for min/max green time constraints, actions are normalized within the range $[-1, 1]$, achieved via a tanh function, and subsequently linearly mapped to actual phase duration as follows:
\begin{equation}
    \tilde{a}_p = \sigma_p^- + \frac{(a_p + 1) \cdot (\sigma_p^+ - \sigma_p^-)}{2},
\end{equation}
where $\tilde{a}_p$ and $a_p$ are the actual and normalized duration time for phase $p$, and $\sigma_p^+, \sigma_p^-$ are the maximum and minimum green time also for $p$. For notational simplicity, we refer to $a$ as the actual duration time in the rest of this paper.

\textbf{Reward} $r$. Let $l$ be the index of each incoming lane, $\mathit{L}$ be the total number of incoming lanes, and $X_l^t$ be the queue length at timestep $t$ for the $l^{th}$ lane. We define the immediate reward as follows:
\begin{equation}
    r_t = -\sum_{l \in \mathit{L}}{\frac{X_l^{t+1}}{\Theta_m}},
\end{equation}
\noindent where $\Theta_m$ is the maximum distance from the stop line that allows vehicle detection. Queue length is often used as a surrogate reward for vehicle delay \cite{wei2021recent}.

\textbf{Sojourn time} $j$. The sojourn time in this context is the time from applying one phase duration (the action) until the next point of applying another phase duration (the next action). The sojourn time is typically the phase duration plus the yellow signal transition time. However, if reservice is applied, the time when the next action is executed will be delayed by the re-service phase, which extends the sojourn time accordingly. 

\subsection{Queue estimation and phase re-service} \label{estimation-reservice}

\textbf{Maximum queue estimation.} First, we give a maximum queue estimation technique for this problem setup, which is slightly altered form of the queue estimation techniques presented in~\cite{yao2019sampled}.

We define the following time quantities related to queue length estimation:
\begin{itemize}
    \item $T^n_g$: The start time when the protected movement in the $n$-th cycle is served with a green signal\footnote{$T$ is the actual simulation time in seconds. In contrast, $t$ is the timestep for the agent's decision-making.}.
    \item $T^{n+1}_r$: The end time of the protected movement in the $n$-th cycle, which is also the $n+1$ cycle's start time.
    \item $T_{re}^{n}$: The time at which whether to re-service is decided for the $(n)$-th cycle.
\end{itemize}

Let $X(T)$ refer to the queue length at time $T$. Additionally we define:
\[
\Delta T^{n} = T_{g}^{n+1} - T^{n}_{re},
\]
as the time between assessing for re-service and the next regular green signal.

At a given $T^{n}_{re}$, $\Delta T^{n}$ could be used to perform queue length estimation, however it is not known in real-time. Instead, let $\Delta \Tilde{T}^{n}$ be a real-time estimate. In this work we estimate $\Delta \Tilde{T}^{n}$ at a given $T^{n}_{re}$ by using the running average of the true $\Delta T^{n}$ from prior cycles. In implementation we use 2 prior cycles.


Let $L_{max}^{n+1}$ be the forecasted maximum queue length in the next cycle, \textit{if no re-service is applied}. We calculate this as follows:
\begin{equation}
    \label{X_max_est}
    L_{max}^{n+1}=v_1^{n}\left( \frac{v_2 \Delta \Tilde{T}^{n} + X(T_{re}^{n})}{v_2-v_1^{n}}\right) + X(T_{re}^{n})\ .
\end{equation}
From $L_{max}^{n+1}$, whether or not to apply phase re-service is decided. The different quantities covered here are shown graphically in Fig.~\ref{fig:shock-wave-simple}.

\textbf{Re-service duration calculation.} First, whether or not to execute the re-service is decided via
\[
L^{n+1}_{max} > \Theta_X,
\]
where $\Theta_X$ is a threshold on queue length. 

Next, we calculate the maximum queue length \textit{if re-service is applied} as follows:
\[
L_{re,max}^{n+1}=\frac{v_2 X(T_{re}^{n})}{v_2-v_1^{n+1}}.
\]

We then calculate the duration of the re-service as follows:
\begin{align}
    &\Delta T_{re}^{n+1} = \notag\\
    &\begin{cases} \label{reservice-duration}
    \text{clip}\left[\sigma_{re}^-, \zeta\frac{L_{re,max}^{n+1}}{v_{3}^{n}}, \sigma_{re}^+\right],& \text{if } X(T_{re}^{n}) < \Theta_X\\
    \sigma_{re}^+,              & \text{if } X(T_{re}^{n}) \geq \Theta_X \\
    0,                          & \text{otherwise}
    \end{cases},
\end{align}

\noindent where $\Delta T_{re}^{n+1}$ is the assigned re-service duration, $\zeta \in (0, 1]$ is a coefficient balancing re-service urgency and overall intersection management, and $\sigma_{re}^-, \sigma_{re}^+$ are the min/max re-service durations. A re-service movement has queues analogous to the right-hand side in Fig. \ref{fig:shock-wave-simple}.





\subsection{Controller training framework} \label{training_framework}


In Algorithm~\ref{pseudocode} our proposed training algorithm is presented. The algorithm joins queue estimation based phase re-service with our SMDP formulation and standard policy optimization techniques. In particular, proximal policy optimization with a generalized advantage estimator is used. Minor alterations to the PPO algorithm were made to adapt the algorithm to the SMDP approach. 



\begin{algorithm} 
\caption{Episodic training process}
\begin{algorithmic}[1] 
\Function{Re-service}{$k_a^{n+1}$,$q_a^{n+1}$} \label{pseudocode}
\State Forecast $L_{max}^{n+1}$ in (\ref{X_max_est})
\State Calculate $\Delta T_{re}^{n+1}$ in (\ref{reservice-duration}), \Return $\Delta T_{re}^{n+1}$
\EndFunction
\Procedure{episode }{episode length $T_{ep}$}
\State Initialize env: $s_0, n \gets 0, T \gets 0, t \gets 0, k_a^n \gets 0, q_a^n \gets 0$
\While{$T < T_{ep}$}
    \State Agent samples $a_t \gets \pi_\theta(\cdot|s_t)$
    \State Env executes $a_t$, returns $s_{t+1},r_t,j_t, n, T \gets T+j_t$
    \If {$T=T_{re}^n$}
        \State Update $k_a^{n+1}, q_a^{n+1}$
        \State Acquire $\Delta T_{re}^{n+1} \gets$ \Call{Re-service }{$k_a^{n+1}, q_a^{n+1}$}
        \If {$\Delta T_{re}^{n+1} > 0$}
            \State Environment executes $a_{re}=\Delta T_{re}^{n+1}$, returns $s_{t+1},r_t,j_t\gets j_t+\Delta T_{re}^{n+1}, T \gets T+\Delta T_{re}^{n+1}$
        \EndIf
    \EndIf
    \State Save $(s_t, a_t, r_t, s_{t+1},j_t)$ in buffer
    \If{buffer is full}
        \State $\pi_\theta$ updated via PPO 
    \EndIf
\EndWhile
\EndProcedure
\end{algorithmic}
\end{algorithm}

\section{Experimental results} \label{experiments}
We present the results of a series of numerical experiments conducted in the traffic microsimulation software SUMO~\cite{krajzewicz2012sumo}. Two different signal control environments are considered, namely a signalized intersection at a freeway ramp and a conventional four-leg intersection. 

Three different ATSC algorithms are compared in both environments. We implement our approach of RL with re-service, as well as two other approaches. They are RL without re-service, and the SOTL algorithm~\cite{cools2013self}. The average vehicle delay, the average number of stops, and total throughput are all measured and compared across a set of different demands. 

\begin{figure}[htbp!]
    \centering
    \subcaptionbox{Experimental intersection 1: A signalized freeway ramp.}[0.45\textwidth]{
        \includegraphics[width=0.35\textwidth]{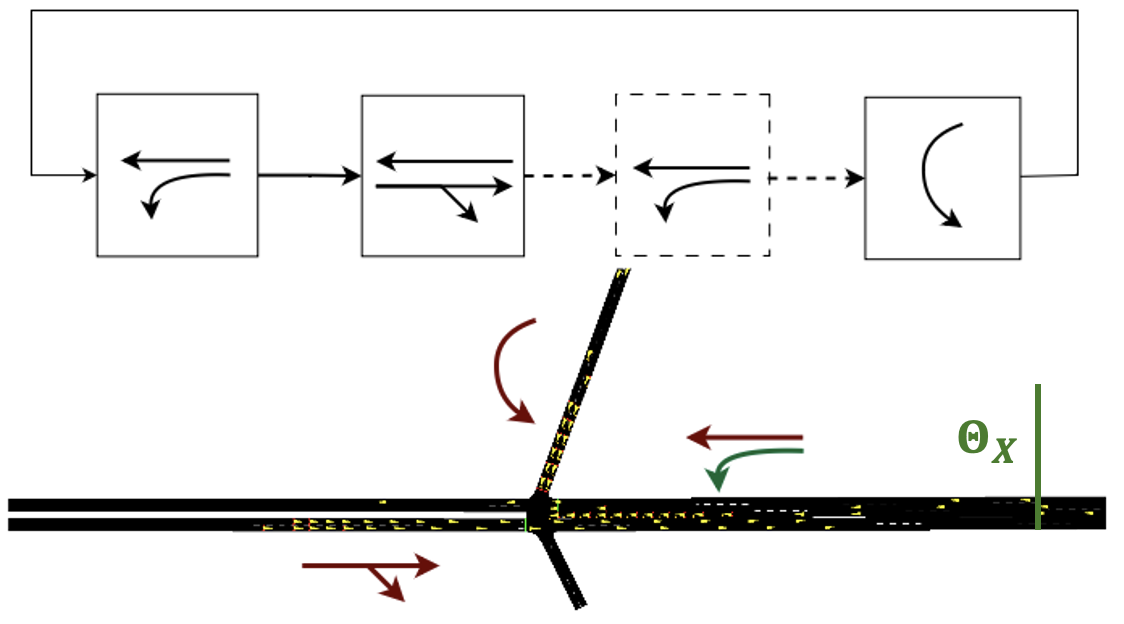}}
    \subcaptionbox{Experimental intersection 2: Four-leg intersection.}[0.45\textwidth]{
        \includegraphics[width=0.35\textwidth]{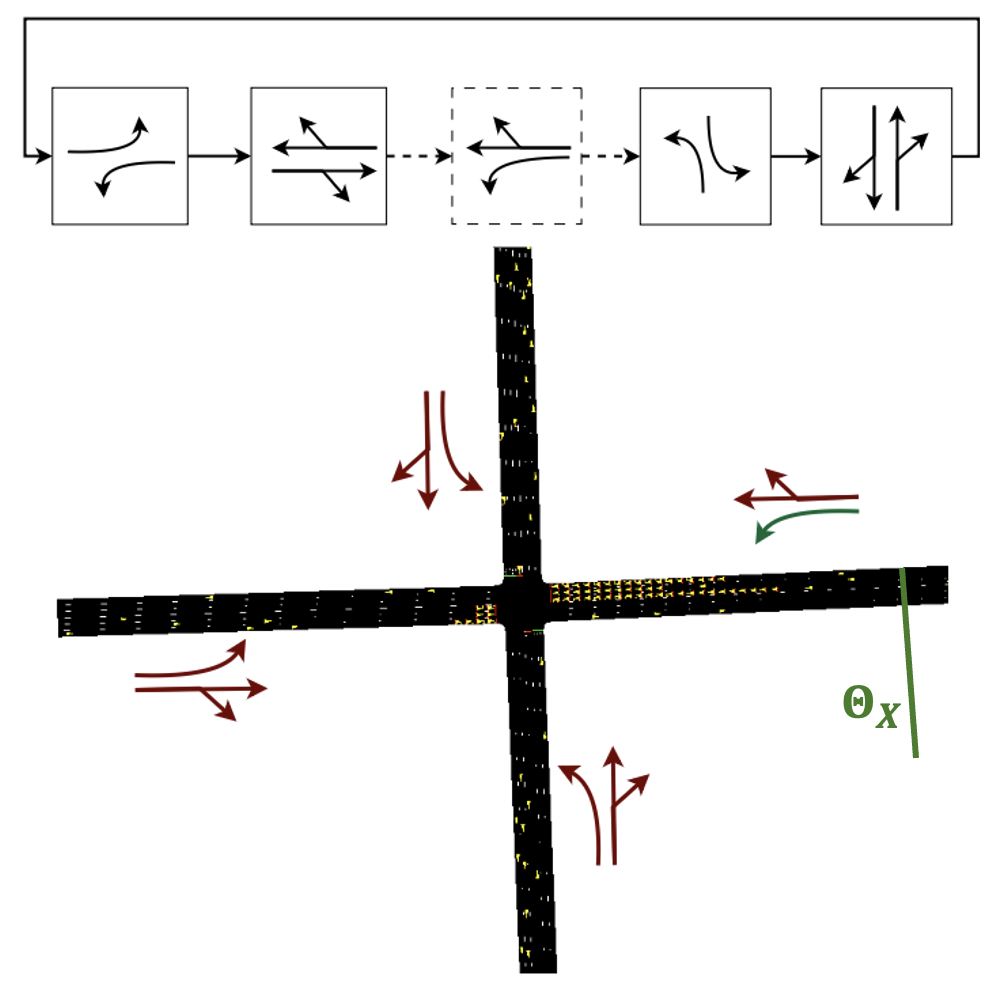}}
    \caption{Experimental scenarios and their phase sequences are shown. Vehicle movements are shown as arrows, protected ones in green and others in red. Regular and re-service phases are boxes with solid and dotted lines.}
    \label{fig:scenario}
\end{figure}

\subsection{Implementation details}
We model the two intersection types in SUMO using the default parameters to control the vehicle dynamics. At each intersection (Fig. \ref{fig:scenario}), we define five time varying demand profiles, which are shown in Appendix A. 

The [min, max] green time constraints for each phase are [5, 30], [5, 40], \textit{[10, 25]}, [5, 45] for the freeway ramp intersection, and [5, 25], [5, 70], \textit{[5, 25]}, [5, 25], [5, 70] for the four-leg intersection, all in seconds. Green time constraints for the re-service phases are in \textit{Italic}. The yellow signal is set at five seconds. The simulation time for all scenarios is one hour. The maximum distance for vehicle detection $\Theta_m$ is set at 250 m, while the threshold for re-service $\Theta_X$ is set at 200 m. The shock wave parameters are estimated from the simulation model parameters as: 
 $k_j=133.3$ veh/km, $k_m=50$ veh/km, and $q_m=1550$ veh/h. 
 

The actor network has a single 128-neuron hidden layer and a tanh activation function for output. Similarly, the critic network also has a hidden layer of 128 neurons but the activation function is ReLU. Both networks are independent, i.e., do not share common neurons. The sampled action value from the actor's stochastic policy is also activated by tanh. The re-service coefficient is set as $\zeta=0.7$. For hyperparameters in PPO, we follow the guidelines in~\cite{andrychowicz2020matters} and set the loss clipping hyperparameter $\epsilon=0.1$, the long-term reward discount factor $\gamma=0.995$, the multi-step weighting factor in advantage estimation $\lambda=0.99$, the learning rate at 2.5e-4, the minibatch size at 256, the update interval of 1200 transitions, and 20 epochs per update. All agents have the same hyperparameters.

\begin{figure}[tbp!]
    \centering
    \includegraphics[width=0.45\textwidth]{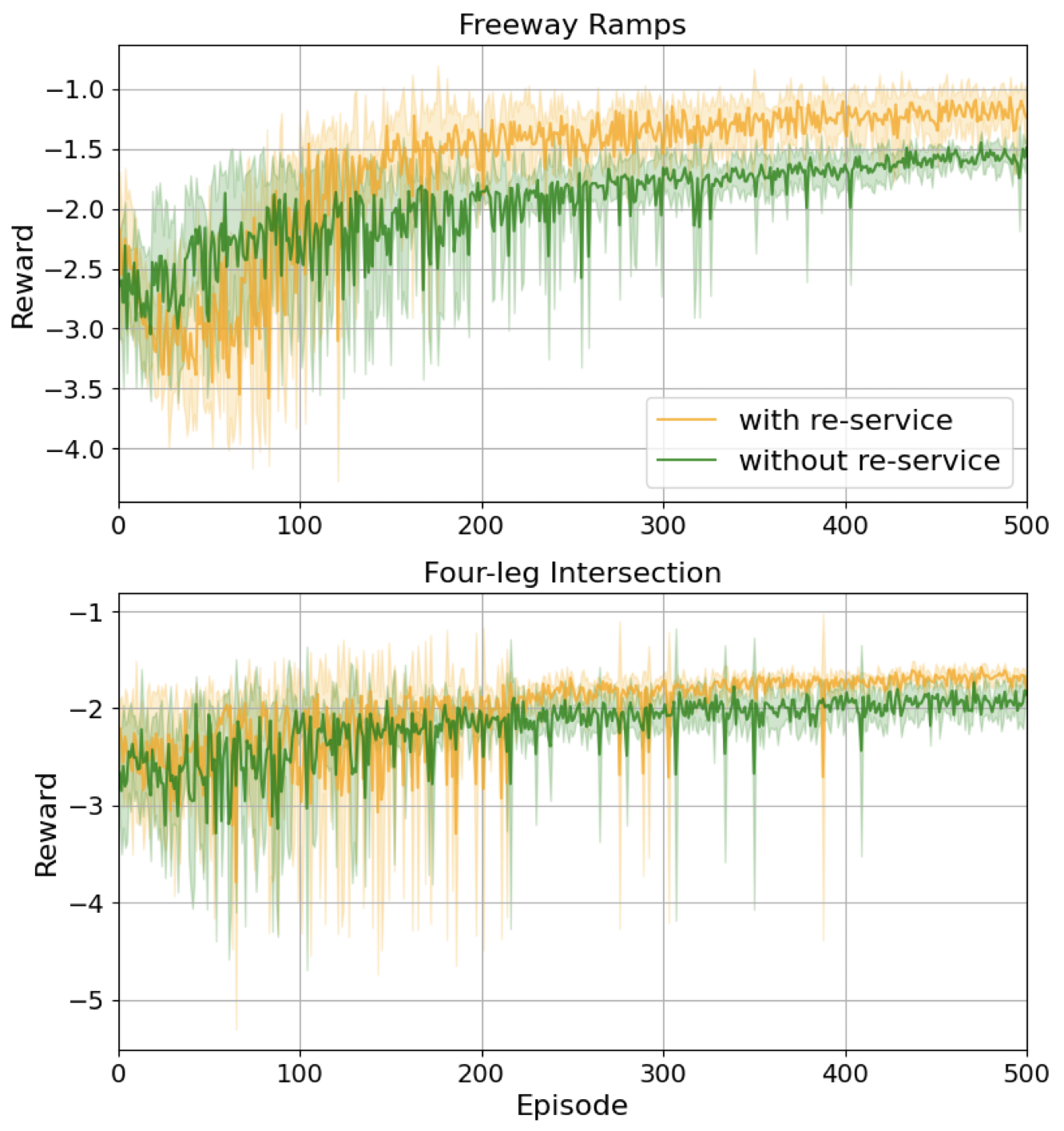}
    \caption{Step-average reward curves for 5 runs. Solid lines are averages and intervals are standard deviations.}
    \label{fig:training}
\end{figure}

\subsection{Training results}
The step-average training results for 500 episodes in two intersections are shown in Fig. \ref{fig:training}. Demand 3 of both intersections are training demand profiles. It shows that adding phase re-service as part of the environmental transition does not significantly affect the speed of convergence, and in both intersections, the re-service can further maximize the rewards. The freeway ramps are more benefited from phase re-service probably due to a large percentage of re-serviced vehicles in the total vehicle demands.

\begin{table*}[htbp]
\centering
\caption{Summary of Test Results (mean, std) in freeway ramps intersection. Percentage of re-service cycles with mean values only (std ignored due to insignificance).}
\label{table:metrics-ramp}
\begin{tabular}{ccccccc}
\toprule
Metric & Algorithm & Demand 1 & Demand 2 & Demand 3 & Demand 4 & Demand 5 \\

 \cmidrule{1-7}
\multirow{3}{*}{Vehicle delay (s)} & with re-service & 34.967, 27.853 & \textbf{42.813}, 36.168 & \textbf{48.393}, 42.29 & \textbf{48.703}, 36.119 & \textbf{43.019}, 35.327\\
 & without re-service & 34.62, 30.933 & 44.579, 47.773 & 57.322, 63.986 & 64.876, 69.702 & 61.497, 65.379\\
  & SOTL & \textbf{33.167}, 35.353 & 57.203, 75.361 & 88.622, 113.092 & 81.864, 102.859 & 70.121, 90.853\\
 \cmidrule{1-7}
\multirow{3}{*}{Number of stops} & with re-service & \textbf{0.681}, 0.508 & \textbf{0.807}, 0.715 & \textbf{0.876}, 0.772 & \textbf{0.897}, 0.711 & \textbf{0.844}, 0.765\\
 & without re-service & 0.715, 0.59 & 0.887, 0.898 & 1.12, 1.223 & 1.213, 1.311 & 1.16, 1.225\\
  & SOTL & 0.822, 0.887 & 1.46, 2.059 & 2.394, 3.025 & 2.181, 2.804 & 1.896, 2.608\\
\cmidrule{1-7}
\% of re-service cycles& with re-service & 6.4 & 15.0 & 23.5 & 45.3 & 26.6\\
\cmidrule{1-7}
\multirow{3}{*}{Throughput (veh/h)} & with re-service & \textbf{1945}, 6.59 & 2352, 6.9 & \textbf{2694}, 8.09 & \textbf{2689}, 10.06 & 2593, 6.84 \\
 & without re-service & 1941, 5.3 & \textbf{2358}, 10.37 & 2690, 7.75 & 2611, 13.59 & \textbf{2594}, 5.94 \\
  & SOTL & 1944, 4.08 & 2314, 13.18 & 2660, 13.36 & 2516, 13.41 & 2510, 13.91 \\
\bottomrule
\end{tabular}
\end{table*}

\begin{table*}[htbp]
\centering
\caption{Summary of Test Results (mean, std) in four-leg intersection. Percentage of re-service cycles with mean values only (std ignored due to insignificance).}
\label{table:metrics-fourleg}
\begin{tabular}{ccccccc}
\toprule

Metric & Algorithm & Demand 1 & Demand 2 & Demand 3 & Demand 4 & Demand 5\\

 \cmidrule{1-7}
\multirow{3}{*}{Vehicle delay (s)} & with re-service & \textbf{67.573}, 51.672 & \textbf{66.355}, 51.913 & \textbf{65.159}, 53.597 & \textbf{66.588}, 52.591 & \textbf{72.759}, 55.823 \\
 & without re-service & 76.605, 77.757 & 72.344, 67.589 & 72.197, 69.815 & 89.74, 128.937 & 80.012, 93.294\\
  & SOTL & 75.689, 80.501 & 80.274, 90.484 & 73.913, 86.785 & 92.121, 149.148 & 83.101, 118.338 \\
 \cmidrule{1-7}
\multirow{3}{*}{Number of stops} & with re-service & \textbf{0.841}, 0.504 & \textbf{0.853}, 0.53 & \textbf{0.83}, 0.509 & \textbf{0.856}, 0.527 & \textbf{0.905}, 0.542\\
 & without re-service & 0.926, 0.671 & 0.885, 0.597 & 0.878, 0.589 & 1.046, 1.068 & 0.994, 0.831\\
  & SOTL & 0.971, 0.758 & 0.97, 0.78 & 0.959, 0.815 & 1.095, 1.266 & 1.084, 1.11\\
\cmidrule{1-7}
\% of re-service cycles& with re-service & 4.8 & 11.0 & 4.0 & 13.7 & 14.1\\
\cmidrule{1-7}
\multirow{3}{*}{Throughput (veh/h)} & with re-service & \textbf{2906}, 12.6 & 2739, 9.3 & \textbf{2782}, 13.3 & \textbf{2613}, 11.9 & \textbf{3130}, 12.5 \\
 & without re-service & 2903, 14.9 & \textbf{2745}, 16.4 & 2771, 11.1 & 2576, 6.3 & 3118, 10.6\\
  & SOTL & 2868, 5.1 & 2701, 3.0 & 2755, 6.4 & 2543, 8.7 & 3116, 7.0 \\
\bottomrule
\end{tabular}
\end{table*}

\begin{figure}[htbp!]
    \centering
    \includegraphics[width=0.4\textwidth]{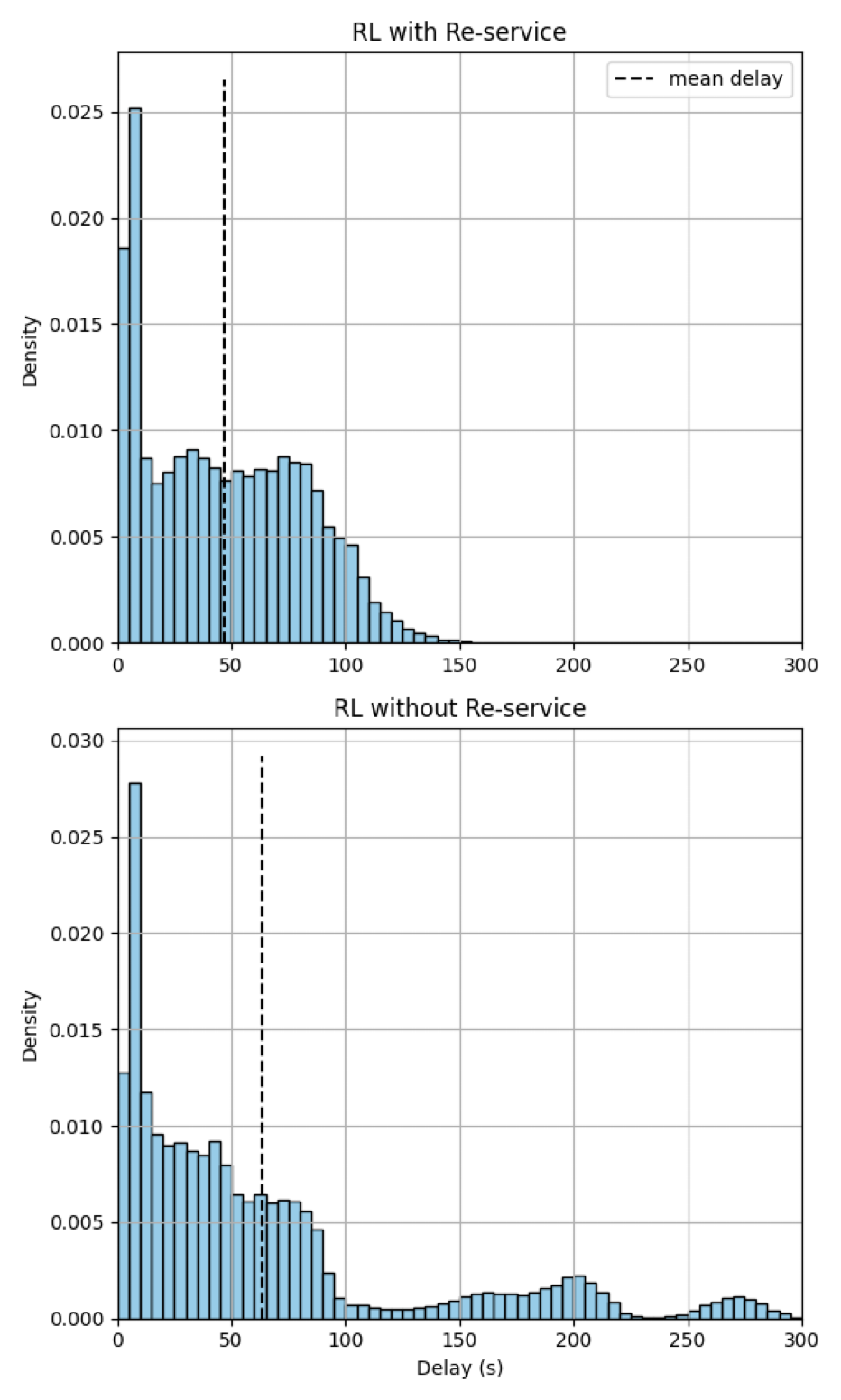}
    \caption{Density histogram of vehicle delays of the RL agent with and without phase re-service in freeway ramps Demand 4 scenario.}
    \label{fig:hist}
\end{figure}

\subsection{Testing performance}
For each intersection, the agent is trained five times with demand profile 3. The best-performing agent in the last episode of the training stage is used for testing. We run each testing scenario (consisting of an intersection and a demand profile) 20 times and report the statistics. Metrics including throughput per hour, vehicle delay per trip, and the number of stops experienced per trip are compared and summarized in Table \ref{table:metrics-ramp} and Table \ref{table:metrics-fourleg}. All metrics are directly calculated within SUMO. 
A baseline non-RL algorithm called SOTL \cite{cools2013self} is evaluated for comparison purposes. SOTL determines phase switching based on the count of arriving vehicles showing excellent performance in previous studies \cite{zheng2019learning}. Below we analyze different traffic performance metrics. 

\textbf{Vehicle delay:} We report the average delay per vehicle. In 9 out of 10 scenarios tested, vehicle delay is significantly reduced with re-service compared to scenarios without re-service and to the non-RL baseline. Numerically, compared to RL without re-service, the proposed approach achieves an improvement in terms of the mean and standard deviation of vehicle delay ranging from -1\% to 29.95\% and 9.95\% to 59.21\%, respectively. SOTL, in contrast, only exhibits a 4.2\% mean vehicle delay improvement with 14.29\% higher variance in Table \ref{table:metrics-ramp} Demand 1. Further, we present the density histogram of vehicle delays for freeway ramps Demand 4 scenario which benefits most from the phase re-service in Fig.~\ref{fig:hist}. RL with re-service method is able to reshape the distribution with a long tail, i.e., significantly delayed vehicles, back to a more centralized one, leading to a lower average delay and smaller standard deviation.

\textbf{Number of stops:} The average number of vehicle stops is reported as a measure of mobility. The phase re-service significantly reduces the average number of stops and the variance of number of stops over all scenarios. The percentage of improvement reaches as high as 26.05\% and 45.77\% in terms of the mean and standard deviation (Table \ref{table:metrics-ramp}, Demand 4 scenario).

\textbf{Percentage of re-service cycles:} This metric calculates the proportion of cycles incorporating phase re-service in each scenario. The results from Table \ref{table:metrics-fourleg} Demand 3 suggest that the re-service cycle as low as 4.0\% can reduce the average vehicle delay and number of stops by 9.75\% and 5.47\%, and moreover, lower the variance of both metrics by 23.23\% and 13.58\%. The highest re-service rate reaches 45.3\% (Table \ref{table:metrics-ramp}, Demand 4 scenario) which contributes to the largest improvements of both vehicle delay and number of stops.

\textbf{Throughput:} The average throughput is also calculated for each scenario. The RL algorithm with and without phase re-service realizes similar performance in maximizing the throughput, which is also comparable to the baseline. 

\subsection{Performance metrics by directions}
We list the trip-level evaluation metrics by vehicle movements of the RL algorithm with and without phase re-service for the scenarios with the lowest and highest re-service penetration. They are summarized in Table \ref{table:direction-metrics}. Through and right-turn movements are grouped together in the four-leg intersection as they utilize the same phases.

As expected, we observe that movements only in regular phases are delayed and stop more since the phase re-service takes additional time. Nevertheless, the additional delay is mild. EE and SE  movements in Table \ref{tab:direction-metrics-ramp-demand-four} are on average more delayed by 5.6s and 13.5s; non-protected movements in Table \ref{tab:direction-metrics-four-leg-demand-3} experience more delays from 3.8s to 14.5s. In contrast, the improvement for the protected movement is substantial: 102.5s less delay and 1.454 fewer stops in Table \ref{tab:direction-metrics-ramp-demand-four}, 79.98s less delay, and 0.45 fewer stops in Table \ref{tab:direction-metrics-four-leg-demand-3}.

\begin{table}[htbp]
\centering
\caption{Trip-level evaluation metric by vehicle movements. \textbf{\textit{Bold and Italic}} font: protected movement.}
\label{table:direction-metrics}
\begin{subtable}{0.52\textwidth}
\caption{Ramp Demand 4 Scenario}
\label{tab:direction-metrics-ramp-demand-four}
\begin{tabular}{cccc}
\toprule
Metric & Movement & with re-service & without re-service\\
\cmidrule{1-4}
\multirow{4}{*}{Vehicle delay (s)} & EE & 55.906, 30.871 & 49.572, 27.026 \\
& WW & 14.108, 12.629 &18.604, 15.085\\
& \textbf{\textit{WS}} & 71.994, 26.755 & 174.5, 65.297\\
& SE & 54.07, 31.034 & 40.508, 26.181\\
\cmidrule{1-4}
\multirow{4}{*}{Number of stops} & EE & 0.907, 0.389 & 0.862, 0.386\\
& WW & 0.344, 0.477 & 0.508, 0.606\\
& \textbf{\textit{WS}} & 1.567, 0.672 & 3.021, 1.635\\
& SE & 0.855, 0.389 & 0.743, 0.44\\
\bottomrule
\end{tabular}
\vspace{5pt}
\end{subtable}
\begin{subtable}{0.52\textwidth}
\caption{Four-leg Demand 3 Scenario}
\label{tab:direction-metrics-four-leg-demand-3}
\begin{tabular}{cccc}
\toprule
Metric & Movement & with re-service & without re-service\\
\cmidrule{1-4}
\multirow{8}{*}{Vehicle delay (s)} & NN\&NE & 47.422, 38.377 & 50.712, 37.564\\
& \textbf{\textit{NW}} & 134.784, 92.833 &214.751, 127.958\\
& WW\&WN & 62.366, 41.288 &55.418, 40.203\\
& WS & 79.247, 50.538 & 83.084, 46.177\\
& EE\&ES & 56.873, 42.269 & 51.937, 38.735\\
& EN & 90.578, 51.432 & 95.351, 54.744\\
& SS\&SW & 46.113, 38.467 &50.441, 37.545\\
& SE & 75.983, 44.467 & 90.613, 46.752\\
\cmidrule{1-4}
\multirow{8}{*}{Number of stops} & NN\&NE & 0.694, 0.473 & 0.761, 0.472\\
& \textbf{\textit{NW}} & 1.382, 0.789 & 1.832, 0.977\\
& WW\&WN & 0.829, 0.428 & 0.754, 0.459\\
& WS & 0.915, 0.358 & 0.908, 0.309\\
& EE\&ES & 0.77, 0.431 & 0.741, 0.442\\
& EN & 0.964, 0.365 & 1.038, 0.393\\
& SS\&SW & 0.684, 0.471 &0.727, 0.452\\
& SE & 0.896, 0.311 & 0.995, 0.363\\
\bottomrule
\end{tabular}
\end{subtable}
\end{table}

\section{Conclusion} \label{conclusion}
In this paper, we propose a method to augment the RL-based ATSC to include temporary phase re-service, aiming to reduce vehicle delays and stops at intersections in high-volume left-turn scenarios. An RL agent determines the duration of the next regular phase, and another rule-based logic incorporating the shock wave theory estimates the queue growth and determines the phase re-service. We formulate the RL problem as SMDP and use PPO to solve it. We test the framework against 2 types of intersections and 10 demand profiles, and demonstrate the general merit of our framework in reducing the vehicle delays and the number of stops overall by up to 29.95\% and 26.05\% of the average and up to 59.21\%
and 45.77\% of the standard deviation.


\bibliographystyle{IEEEtran}
\bibliography{references}

\appendix
\subsection{Demand profiles}\label{sec:demand_profiles}
\begin{figure}[!t]
    \centering
    \begin{subfigure}[b]{\columnwidth} 
        \includegraphics[width=\linewidth]{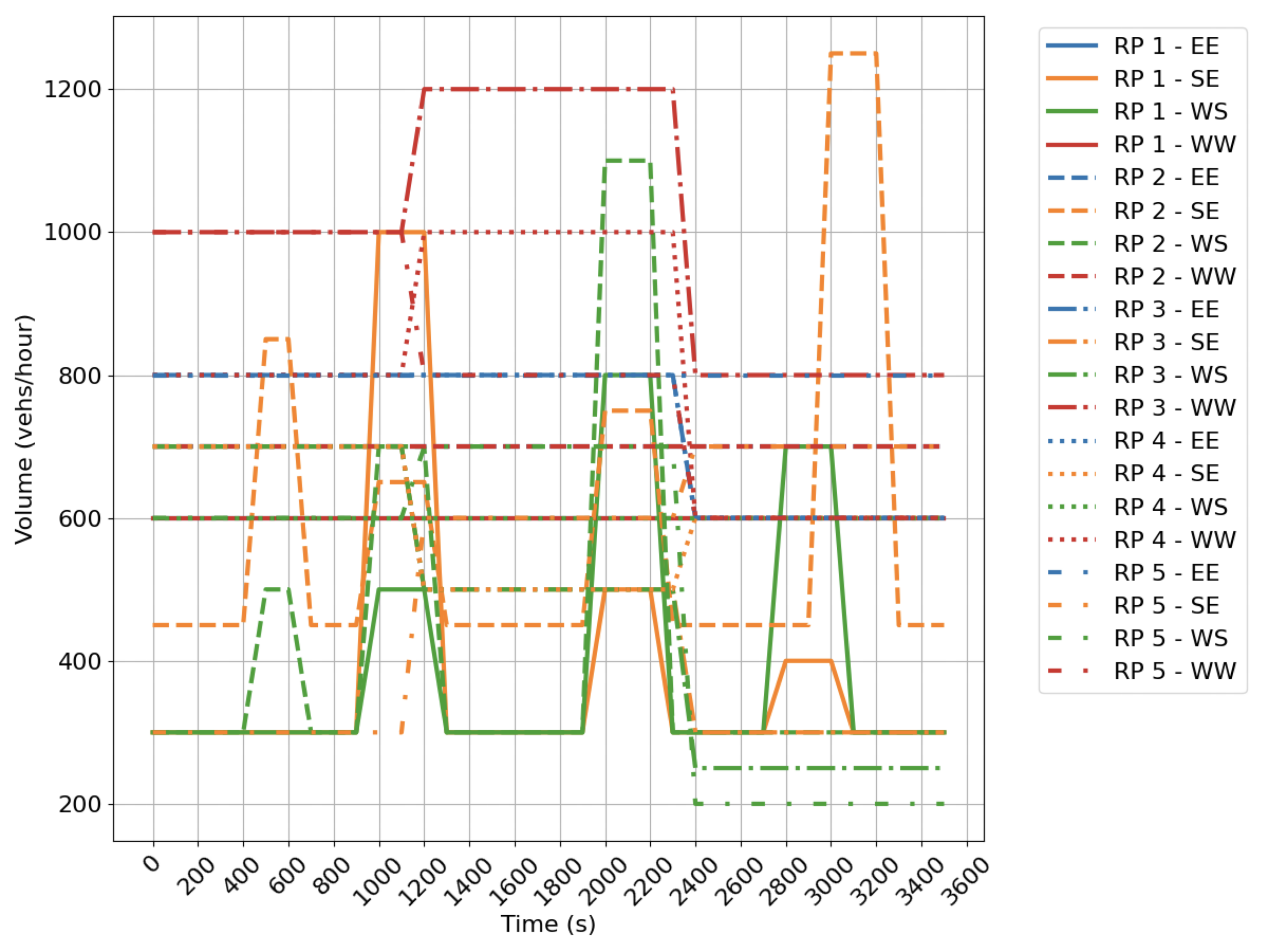} 
        \caption{RP 1-5 scenarios}
        \label{fig:demand-rp}
    \end{subfigure}

    \begin{subfigure}[b]{\columnwidth}
        \includegraphics[width=\linewidth]{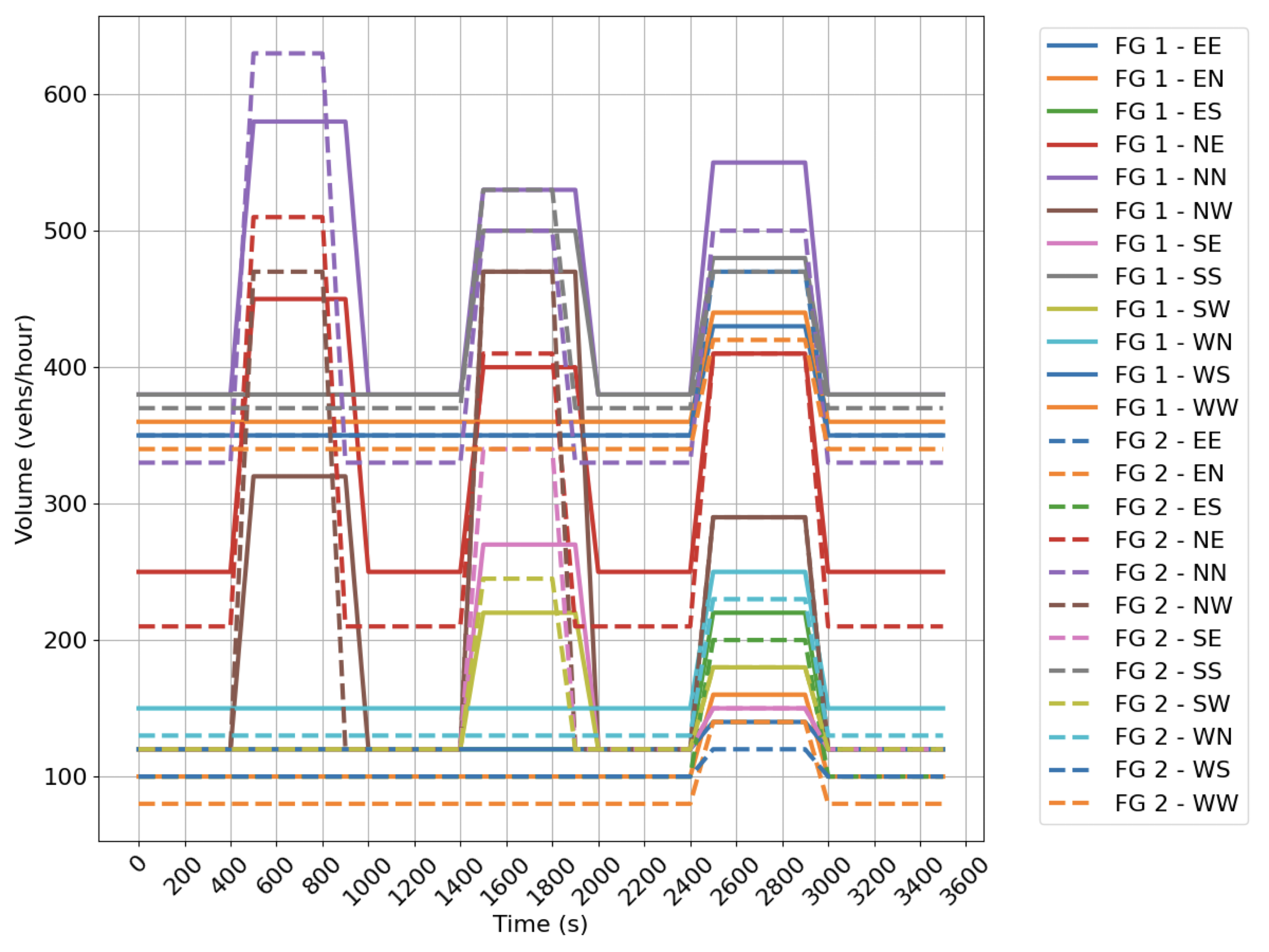}
        \caption{FG 1-2 scenarios}
        \label{fig:demand-fg1}
    \end{subfigure}

    \begin{subfigure}[b]{\columnwidth}
        \includegraphics[width=\linewidth]{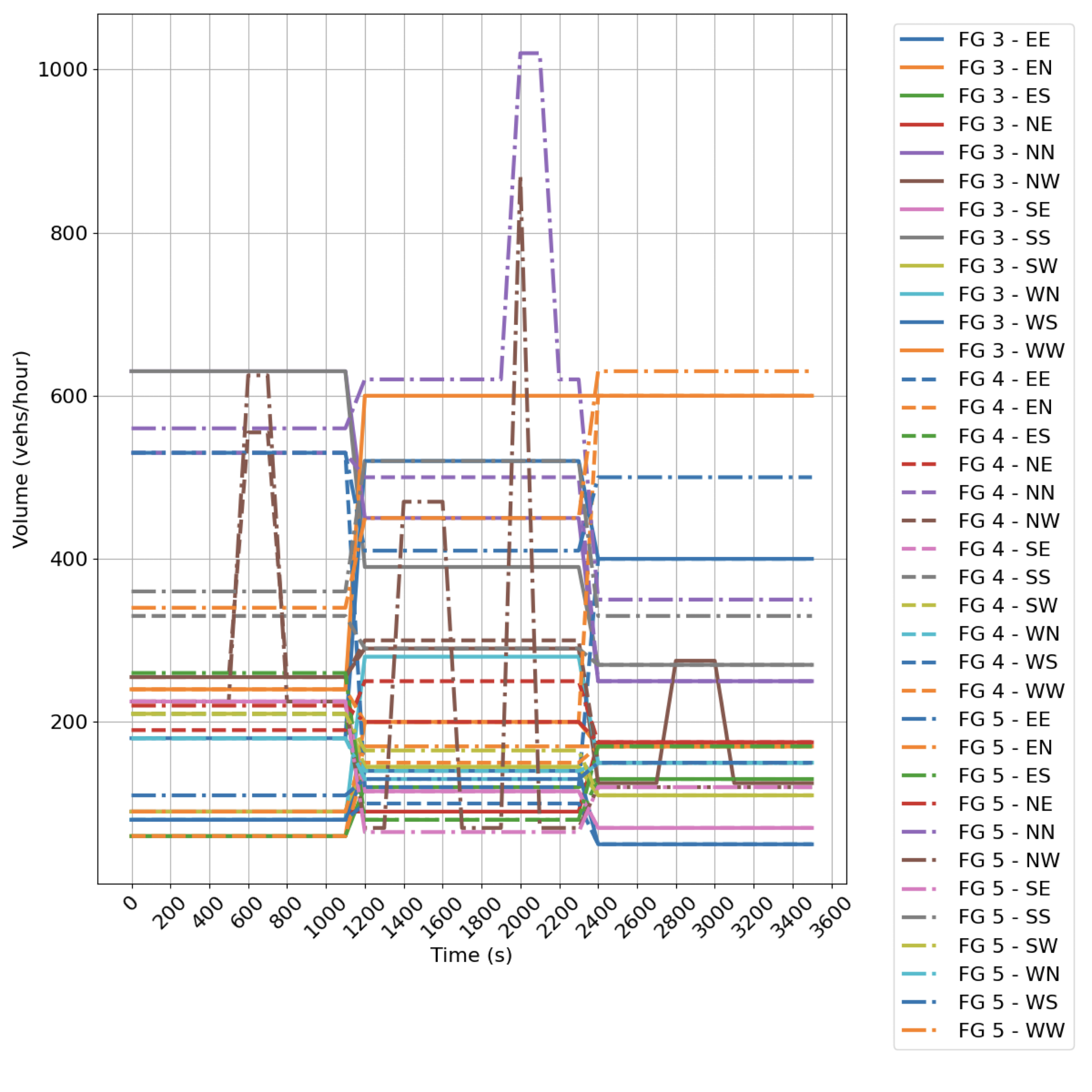}
        \caption{FG 3-5 scenarios}
        \label{fig:demand-fg2}
    \end{subfigure}
    
    \caption{Traffic demands by moving directions for all scenarios.}
    \label{fig:demands}
\end{figure}

 We visualize the time varying demand profiles by intersection movement in Fig.~\ref{fig:demands}. Fig.~\ref{fig:demand-rp} provides the ramp (RP) flows for five scenarios. Fig~\ref{fig:demand-fg1} and \ref{fig:demand-fg2} show the demand profiles for the four leg intersection (FG). 

\end{document}